# Output Width Signal Control In Asynchronous Digital Systems Using Monostabil Circuits

**Mihai Timiş**

"Gh. Asachi" Technical University of Iaşi
Automatic Control and Computer Science and Engineering Faculty
Computer Science and Engineering Department

**ABSTRACT.** In present paper, I propose a method for resolving the timing delays for output signals from an asynchronous sequential system. It will be used an example of an asynchronous sequential system that will set up an output signal when an input signal will be set up. The width of the output signal depends on the input signal width, and in this case it is very short. There are many synthesis methods, like using a RC group system, a monostabil system in design of the asynchronous digital system or using an exetrnal clock signal, CK. In this paper will be used a monostabil circuit.
**KEYWORDS:** FSM Graph, FSM Transition Table, Latch D, Local Clock, Monostable Circuit ,External clock - CK, Logic Gates, RC Group, Digital Systems.

## 1. Introduction

I will porpose an asynchronous digital system who will generate an output signal, named Z, drive by an logic signal named H (high), figure 1.

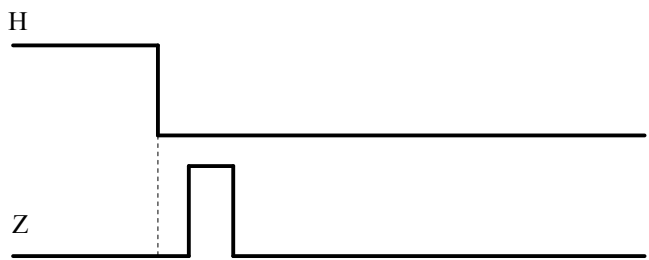

*Fig. 1*

The fluence state graph is illustrated in figure 2.

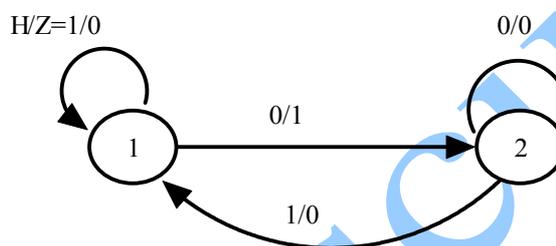

*Fig. 2*

Based on figure 1,2 will be deducted the fluence state table, transition/output matrix, figure 3, 4.

| H \ $Q_n$ | 0   | 1   |
|-----------|-----|-----|
| 1         | 2/1 | 1/0 |
| 2         | 2/0 | 1/1 |

*Fig. 3*

244

| $Q_n$ | y |
|---|---|
| 1 | 0 |
| 2 | 1 |

| $H_n$ \ $y_n$ | $Y_{n+1}$ | | $z_n$ | |
|---|---|---|---|---|
|  | 0 | 1 | 0 | |
| 0 | 1 | 0 | 1 | 0 |
| 1 | 1 | 0 | 0 | 0 |

*Fig. 4*

The asynchronous digital system equations are:

$y_{n+1} = \overline{H_n}$

$z_n = \overline{y_n} \cdot \overline{H_n}$

The timing diagram is illustrating in figure 5.

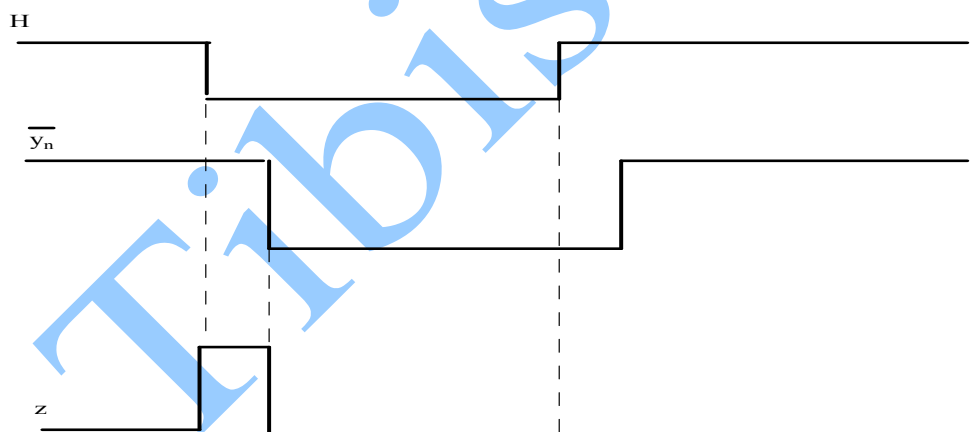

*Fig. 5*

## 2. Method Description

The Z output signal activation will command a monostabil circuit which will generate a M signal. The period of the M signal will depend of the RC constant, figure 6.

245

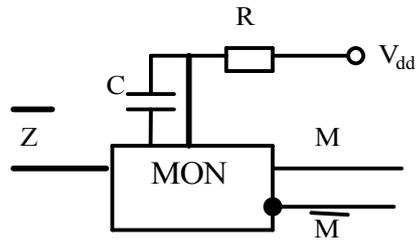

**Fig. 6**

The fluence graph will be modified with those 2 input variables H, M like in figure 7.

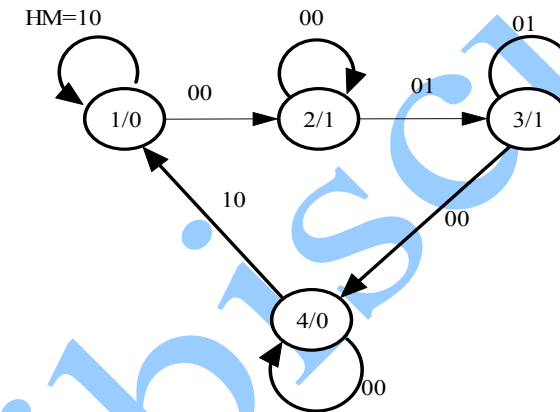

**Fig. 7**

The period of the H signal is too great than the period of the M signal. The fluence table, matrix transition table and output transition table are described in figure 8.

| HM (Q/Z)$_n$ | 00 | 01 | 11 | 10 | HM (y$_1$y$_0$)$_n$ | (y$_1$y$_0$)$_{n+1}$ | | | | z$_n$ | | | |
|---|---|---|---|---|---|---|---|---|---|---|---|---|---|
| | | | | | | 00 | 01 | 11 | 10 | 00 | 01 | 11 | 10 |
| 1/0 | 2 | - | - | 1 | 0 0 | 01 | -- | -- | 00 | 0 | 0 | 0 | 0 |
| 2/1 | 2 | 3 | - | - | 0 1 | 01 | 11 | -- | -- | 1 | 1 | 1 | 1 |
| 3/1 | 4 | 3 | - | - | 1 1 | 10 | 11 | -- | -- | 1 | 1 | 1 | 1 |
| 4/0 | 4 | - | - | 1 | 1 0 | 10 | -- | -- | 00 | 0 | 0 | 0 | 0 |

**Fig. 8**

From figure 8 we will deduct the next equations:

$$\begin{cases} y_{1,n+1} = M \cdot (\overline{H} + y_1) \\ y_{0,n+1} = M \cdot (\overline{H} + \overline{y_1}) \\ z_n = y_0 \end{cases}$$

The architecure of the asinchronous sequential system is ilustrated in figure 9, (Mealy system), $C_d = \sum_{0}^{q-1} z_i$.

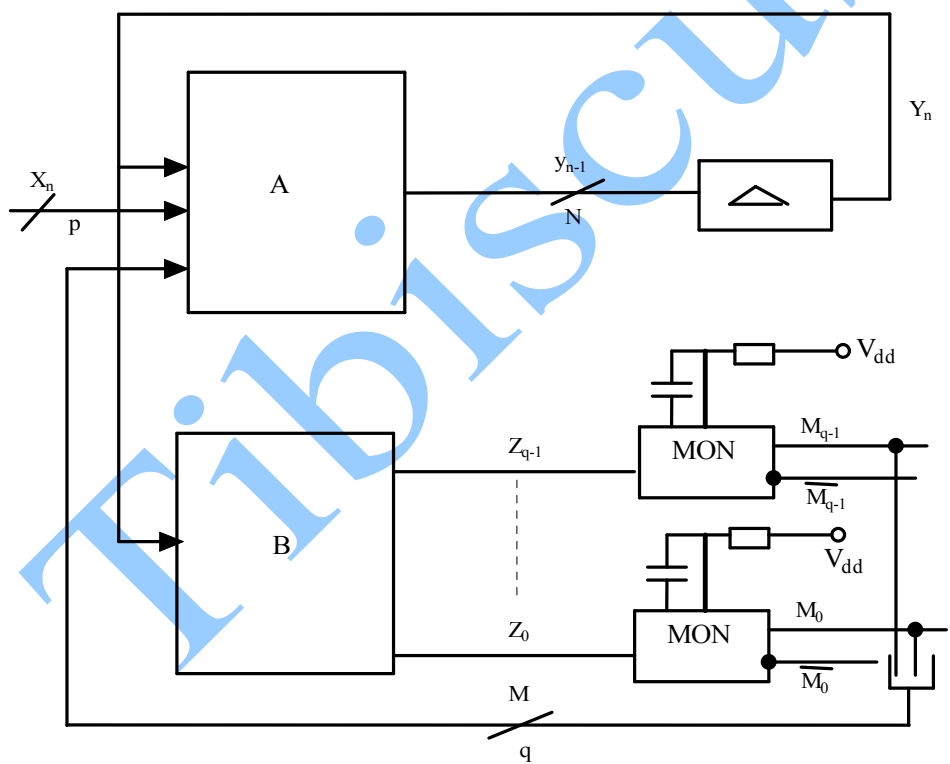

*Fig. 9*

247

**Conclusions**

We will propose two methods for enlarge the width of the Z output signal:
- Using a RC circuit (the Z output signal will drive a monostabil who will generate a signal, proportional with the RC constant).
- Using an external clock (CK) signal with ½ full fill factor. This signal will be considered only when the output signals will be generated.